\documentclass[aps,preprint,nofootinbib,showpacs]{revtex4}
\usepackage{epsf,graphics,graphicx}
\usepackage{amsmath}
\usepackage{amssymb,latexsym,mathrsfs}

\def\bea{\begin{eqnarray}}
\def\eea{\end{eqnarray}}
\def\ba{\begin{array}}
\def\ea{\end{array}}

\def\ket{\rangle}

\def\beq{\begin{equation}}
\def\eeq{\end{equation}}

\begin{document}

\DeclareRobustCommand{\baselinestretch{2.2}}

\title{Dynamics of the geometric phase in the adiabatic limit of a quench induced quantum phase transition }


\author{B. Basu}
\footnote{banasri@isical.ac.in, Fax:+91(033)2577-3026}
\affiliation{Physics and Applied Mathematics Unit, Indian
Statistical Institute, Kolkata 700108, India}
\begin{abstract}


The geometric phase associated with a many body ground state
exhibits a signature of quantum phase transition. In this context,
we have studied the behavior of the geometric phase during a linear
quench caused by a gradual turning off of the magnetic field
interacting with  a spin chain.
\end{abstract}

 \pacs{64.70.Tg, 03.65.Vf\\
 keywords-~~geometric phase, spin model, quantum phase transition, slowly varying magnetic field}
\maketitle

Geometric phases have been associated with a variety of condensed matter and solid state phenomena
\cite{thou,resta,hatsugai,bp,bps,bb} since its inception \cite{berry}. Besides, various theoretical investigations, geomteric phases have been experimentally tested in various cases, e.g. with photons
\cite{p1,p2,p3}, with neutrons \cite{n1,n2} and with atoms \cite{a1}. The generation of a geometric phase (GP) is a witness
of a singular point in the energy spectrum that arises in all non-trivial geometric evolutions. In this respect,
the connection of geometric phase with quantum phase transition (QPT) has been explored very recently
\cite{car,zhu,hamma}. There has been an increasing interest to investigate the importance of GP as a witness of QPT for various many body systems \cite{chen,yi,cui,nesterov,shu,oh,erik}. 
In essence, the small variations of external parameters resulting to
the abrupt changes on the macroscopic behavior of a system is
described by QPT. These critical changes are due to the presence of
degeneracies in the energy spectrum and are characterized by long
range quantum correlations. The geometric phase can be used as a
tool to probe QPT in many body systems.

 Thermal phase transitions occur when the strength of the thermal fluctuations equals a certain threshold and the
 character of the stable phase changes. In contrast to that, zero temperature phase transitions, such as quantum phase
 transitions \cite{ss}, denote the crossover of different ground states at a certain critical value of some external
 parameter, where quantum fluctuations play a dominant role. Since response times typically diverge in the vicinity of
 the critical point, sweeping through the phase transition with a finite velocity leads to a breakdown of adiabatic
 condition and generate interesting dynamical (non-equilibrium ) effects.
In the case of thermal phase transitions, the Kibble-Zurek (KZ) mechanism \cite{kib,zur} explains the formation of
defects via rapid cooling. This idea of  defect
formation in second order phase transition has been extended to zero
temperature quantum phase transition (QPT)\cite{zur1,dziar} in spin
models. The KZ mechanism, explored in the
one dimensional transverse Ising model helped to estimate the
density of defects produced in QPT in the system\cite{zur1}. It is
predicted that the density of kinks scales as
$\tau_q^{-\frac{1}{2}}$, $\tau_q$ being the quench time. For an adiabatic transition, i.e. for large $\tau_q$,
smaller number of kinks are produced whereas for a fast transition, more number of defects are generated.

From the existing literature, one may note that there is an increasing interest to investigate the role of geometric phase in detecting QPT for many body systems from the geometrical perspective \cite{car,zhu,hamma,chen,yi,cui,nesterov} and study the criticality of the system. 
The initial work \cite{car} showing that the geometric phase can be exploited as a tool to detect regions of criticality without having to undergo a quantum phase transition, was followed by subsequent works. 
It was shown \cite{zhu}that the geometric phase of the ground state obeys scaling behavior in the vicinity of a quantum phase transition  and also the geometric phase can be considered as a topological test to reveal QPT \cite{hamma}. In connection with QPT, there is a growing attention of studying the geoemtric phase of the ground state of various spin systems \cite{chen,yi,cui,nesterov}. Very recently, the study of geometric phase of the ground state for a complicated inhomogeneous period-two anisotropic XY model in a transverse field showed that \cite{shu} there may exist more than one QPT point at some parameter regions and these transition points correspond to the divergence or extremum properties of the Berry curvature. 

These results motivated us to see how the geometric phase of the ground state of a spin chain behaves when the external magnetic field is varied. In this context, we have adopted the fascinating approach of studying the dynamics of a quench induced quantum phase transition, and obtain the dynamics of the geometric phase, which is marked as an indicator of QPT.
Specifically, we have considered a spin chain with XY type of interaction in a slowly varying time dependent magnetic field and have studied the behavior of the geometric phase during a linear quench caused by a gradual turning off of the magnetic field interacting with  a spin chain. 
In the adiabatic limit, we have derived  the instantaneous geometric phases of the corresponding quantum states, and showed that the  geometric phase is dependent  on the quench time. 


 We start with the review of the geometric phase of a spin $1/2$ particle in a varying magnetic field. The Hamiltonian of a single spin $1/2$ system in the presence of an external time dependent magnetic field is given by
 \begin{equation}\label{h1}
 H(t)=\frac{1}{2}{\bf B}(t). {\vec{\sigma}}
 \end{equation}
 where ${\vec{\sigma}}=(\sigma_x,\sigma_y, \sigma_z)$ are the Pauli matrices and ${\bf{B}}(t)=B_0{\bf n}(\theta(t))$ with
  the unit vector ${\bf n}=(\sin\theta \cos\phi,\sin\theta\sin\phi,\cos\theta)$. If the external magnetic field is varied  adiabatically the instantaneous energy eigenstates follow the directions of ${\bf n}$ and can be expressed as
  \begin{equation}\label{art}
\begin{array}{ccc}
    \displaystyle{|\uparrow_n;t>}&=&\displaystyle{\cos \frac{\theta}{2} |\uparrow_z> +~ e^{i\phi} \sin
    \frac{\theta}{2}|\downarrow_z> }\\
    &&\\
    \displaystyle{|\downarrow_n;t>}&=&\displaystyle{\sin \frac{\theta}{2} |\uparrow_z> -~ e^{i\phi}\cos
    \frac{\theta}{2}|\downarrow_z>}
\end{array}
\end{equation}
where $|\uparrow_z>,|\downarrow_z>$ are the eigenstates of the
$\sigma_z$ operator.

For a cyclic time evolution i.e. for ${\bf{B}}(T)={\bf{B}}(0)$, apart from the dynamical phase, the  eigenstates acquire a geometric phase also and we can write,
\begin{equation}
|\uparrow_{n(T)}>=e^{i\delta} e^{i\gamma_B}|\uparrow_{n(0)}>
\end{equation}
where the dynamical phase $\delta=\int_{0}^T B_0(t) dt$ and the geometric phase (GP)
$\gamma_B=\oint{\bf A}^\uparrow \cdot d{\vec{\lambda}}$;\\  ${\vec{\lambda}}$ is the set of control parameters and
 ${\bf A}^\uparrow=i< \uparrow_n|\nabla_\lambda|\uparrow_n>$ is the so called Berry connection. In our case,
 ${\vec{\lambda}}=(\theta,\phi)$ and a straight forward calculation shows that
 \begin{equation}
 \gamma_{\uparrow}=-\gamma_{\downarrow}=\pi(1- \cos\theta)
 \end{equation}
 We should note here that though the eigenenergies depend on $B_0(t)$ the eigenstates depend on ${\bf{n}}(t)$ only.
 The GP is proportional to the solid angle subtended by ${\bf{B}}$ with respect to the degeneracy ${\bf{B}}=0$.
It may also be noted that fluctuations in the external magnetic field will obviously induce fluctuation in the geometric phase
($\gamma_{\uparrow}$ or $\gamma_{\downarrow})$ of the corresponding spin  through fluctuation in $\cos\theta$
\cite{palma}. The gradual slowing down of the magnetic field will induce an effect in the angle $\theta$ which will
subsequently affect the geometric phase of  the corresponding state. In the present communication, we have adopted this
idea in a spin system with XY type of interaction and investigated the dynamics of  geometric phase when the system is
 evolved by a slowly varying time dependent magnetic field.

Since XY model is exactly solvable and presents a rich structure, we
extend our analysis for a chain of spin $\frac{1}{2}$ particles with
XY type of interactions to study the dynamics of geometric phase for
an adiabatic evolution. This is a one dimensional model with nearest
neighbor spin-spin interaction and the external magnetic field is
allowed to orient along the z--direction. The Hamiltonian of this
model is given by
 \begin{equation}\label{h1}
H=\sum_{-M}^{+M} \left( \frac{1+\alpha}{2}\sigma_{i}^x \sigma_{i+1}^x~+ \frac{1-\alpha}{2}\sigma_{i}^y \sigma_{i+1}^y+ \frac{B}{2}\sigma_i^z\right)
 \end{equation}
 where $\sigma_i$'s are the standard Pauli matrices at site $i$, $\alpha$ is the anisotropy parameter,  $N=2M+1$ denotes
 the number of sites, $B$ is the strength of the external magnetic
 field.  We assume
 periodic boundary condition.

 In this model, the geometric phase (GP) of the ground state is evaluated by applying a rotation of $\phi$ around the
 $Z$-axis in a closed circuit to each spin \cite{car, car1}. A new set of Hamiltonians $H_\phi$ is constructed from the Hamiltonian
 (\ref{h1}) as
 \begin{equation}\label{h2}
 H_\phi=U(\phi)~H~U^\dagger(\phi)
 \end{equation}
 where
 \begin{equation}\label{g}
 U(\phi)=\prod_{j=-M}^{+M} \exp(i\phi\sigma_j^z/2)
 \end{equation}
 and $\sigma_j^z$ is the $z$ component of the standard Pauli matrix at site $j$.
 The family of Hamiltonians generated by varying $\phi$ has the same energy spectrum as the initial Hamiltonian and $H(\phi)$ is $\pi$-periodic in $\phi$
 With the help of standard Jordan-Wigner transformations, which makes the spins to one dimensional spinless fermions via the relation $a_j=\left( \prod_{i<j}\sigma_i^z \right) \sigma_j^\dagger$
 and then using the Fourier transforms of the fermionic operator\\
$d_k=\frac{1}{\sqrt{N}}\sum_j ~a_j\exp\left( \frac{-2\pi j_k}{N}\right)~~~{\rm{with}}~~~ k=-M,...+M$
 the Hamiltonian $H_\phi$ can be diagonalized by transforming the fermionic operators in momentum space and then using
 Bogoliubov transformation.

The ground state  $|g\ket$  of the system is expressed as
 \begin{equation}\label{g}
 |g\ket=\prod_{k>0}(\cos\frac{\theta_k}{2}|0\ket_k|0\ket_{-k} -i\exp(2i\phi)\sin\frac{\theta_k}{2}|1\ket_k|1\ket_{-k}
 \end{equation}
 where $|0\ket_k$ and $|1\ket_k$ are the vacuum and single fermionic excitation of the $k$-th momentum mode respectively. The angle $\theta_k$ is given by
\begin{equation}
\cos\theta_k=\frac{\cos k-B}{\Lambda_k}
 \end{equation}
 and
 \begin{equation}
\displaystyle{ \Lambda_k=\sqrt{(\cos k-B)^2+\alpha^2\sin^2 k}}
 \end{equation}
 is the energy gap above the ground state.
 The ground state is a direct product of $N$ spins, each lying in the two-dimensional Hilbert space spanned by $|0\ket_k
 |0\ket_{-k}$ and $|1\ket_k |1\ket_{-k}$. For each value of $k$, the state in each of the two dimensional Hilbert space can be represented as a Bloch vector with coordinates $(2\phi,\theta_k)$. The overall phase is given by the sum of the individual phases.
 The pseudomomenta $k$ take $half$ $integer$ values:
\begin{equation}
k=\pm\frac{1}{2}\frac{2\pi}{N},.....,\pm\frac{N-1}{2}\frac{2\pi}{N}
\end{equation}
The direct calculation shows that the geometric phase for the $kth$  mode, which represents the area in the parameter space enclosed by the loop determined by ($2\phi,\theta_k)$ is given by
 \begin{equation}\label{ph}
 \Gamma_k=\pi(1-\cos\theta_k)
 \end{equation}
 The geometric phase of the state $|g\ket$ is given by
 \begin{equation}
 \Gamma_g=\sum_{k}~\pi(1-\cos\theta_k)
 \end{equation}

 For an adiabatic evolution, if the initial state is an eigenstate, the evolved state remains in the eigenstate.  So 
 we may now derive the instantaneous geometric phases of this system due to a gradually decreasing magnetic field.
Let us  explore the situation when  the system (\ref{h1}) is driven adiabatically (slow transition ) by a  time dependent magnetic field $B(t)$ such that
\begin{equation}\label{quench}
 B(t<0)=-\frac{t}{\tau_q}
 \end{equation}
 $B(t)$, driving the transition, is assumed to be linear with an adjustable time parameter $\tau_q$.

 Let the system be initially at time $ t(<0)<<\tau_q$ such that $B(t)>>1$.
 The instantaneous ground state at any instant $t$
is given by
\begin{equation}\label{g}
 |\psi_0(t)\ket=\prod_{k}(\cos\frac{\theta_k(t)}{2}|0\ket_k|0\ket_{-k} -i\exp(2i\phi)\sin\frac{\theta_k(t)}{2}|1\ket_k|1\ket_{-k}
 \end{equation}
We now use eqn. (12) and (9) to derive
 the geometric phase of the $k^{th}$ mode  which yields
 \begin{equation}
\Gamma_k(t)=\pi\left(1-\frac{\cos k +\frac{t}{\tau_q}}{\sqrt{(\cos k +\frac{t}{\tau_q})^2+\alpha^2\sin^2 k}}\right)
\end{equation}
The variation of the geometric phase $\Gamma_k$ with time $t$, for a
fixed $k$ and $\alpha=0.2$ and for different values of the
adjustable parameter $\tau_q$, the quench time, is shown in Figure 1. \\

The geometric phase for an isotropic system with $\alpha=0$ and
quantum Ising model with $\alpha=1$ may now be easily obtained.
\begin{eqnarray}
 \rm{For}~ \alpha = 0, ~~~~~~~~~~~~~~~~~~~~~~~~~~~~~~~~~~\Gamma_k(t)& = & 0  \nonumber \\
 \rm{and}~ \rm{for}~ \alpha = 1,~~~~~~~~~~~~~~~~~~~~~~~~~~~~
 \Gamma_k(t)& =& \pi \left( 1-\frac{\cos k
 +\frac{t}{\tau_q}}{\sqrt{1+\frac{t^2}{\tau_q^2}+2\frac{t}{\tau_q}\cos k}}\right)
\end{eqnarray}

For a system of size $N$ the total geometric phase for the initial state is,
 \begin{equation}
\Gamma_{initial}=\sum_k \Gamma_{k}
\end{equation}
The magnetic field is gradually decreased by adjusting $\tau_q$ and the critical point is attained at the instant $t=-\tau_q$ with $B=1$. Then the geometric phase for the k th mode is
\begin{eqnarray}
 \Gamma_k(t=-\tau_q)&=&\pi \left(1-\frac{\cos k+1}{\sqrt{(\cos k +1)^2+\alpha^2\sin^2 k}}\right)\\
 & = & 0~~~~~~~~~~~~~~~~~~~~~~~~~~~~~~~~~~~~~~~~~~~~~~~\rm{for}~~~~~~~\alpha=0 \\
 & = & \pi \left(1-\sqrt{\frac 12 (\cos k +1)}\right)~~~~~~~~~~~~~~~~~\rm{for}~~~~~~~\alpha=1
 \end{eqnarray}
 At the critical point the total geometric phase is
\begin{eqnarray}
\Gamma_{critical}&=&\sum_k \Gamma_{k}(t= -\tau_q
)
\end{eqnarray}
Finally, at $t=0$, when the magnetic field is gradually turned off, the situation is a bit different.
The configuration of the final state will depend on the number of kinks generated in the system due to phase transition at or near  $t=-\tau_q$ and as such it will depend on the quench time $\tau_q$ \cite{zur}.
The number of kinks is the number of quasiparticles excited at $B=0$ and is given by
\begin{equation}
{\cal{N}}={\sum_k }p_k
\end{equation}
where $p_k$, the excitation probability (for the slow transition) is given by the Landau Zener formula \cite{zener}
\begin{equation}
 p_k \approx \exp{ (-2\pi \tau_q k^2)}
\end{equation}
As different pairs of quasiparticles ($k,-k$) evolve independently, for large values of $\tau_q$, it is likely that only one pair of quasiparticles with momenta $(k_0,-k_0)$ will be excited where $k_0(=\frac{\pi}{N})$ corresponds to the minimum value of the energy $\Lambda_k$.
Thus the condition for adiabatic transition in a finite chain is given by,
\begin{equation}
\tau_q >> \frac{N^2}{2 \pi^3}
\end{equation}
Hence, well in the adiabatic regime, the
final state at $t=0$  is given by
\begin{equation}\label{final}
 |\psi_{final}\ket =|1\ket_{k_0}|0\ket_{-k_0}\nonumber \\
 \prod_{k,k\neq\pm k_0}
 (\cos\frac{\theta_k}{2}|0\ket_k|0\ket_{-k} -i\exp(2i\phi)\sin\frac{\theta_k}{2}|1\ket_k|1\ket_{-k}
 \end{equation}
 This state is similar to direct product of only $N-1$ spins oriented along $(2\phi, \theta_k)$ where the state of the spin corresponding to momentum $k_0$ does not contribute to the geometric phase.
The total geometric phase of this state is given by
\begin{equation}
\Gamma_{final}(t=0)=\sum_{k, k\neq \pm k_0}\pi (1-\cos\theta_k)
\end{equation}
where
\begin{eqnarray}
\cos \theta_k &=& \frac{\cos k}{\sqrt{\cos^2 k + \alpha^2
\sin^2 k}}\\
|\cos\theta_k|& =&  1 ~~~~~~~~~~~~~~~~~~~~~~~~~\rm{for}~~~~~~\alpha=0   \\
|\cos\theta_k|&=& \cos k
~~~~~~~~~~\rm{for}~~~~~\alpha=1
\end{eqnarray}
For $\alpha=0$, i.e. for an isotropic system, the
geometric phase  of the final state is given by
\begin{equation}
\Gamma_{final}(t=0) \approx 0 
\end{equation}
and for a system with $\alpha=1$, {\it i.e.} for the well known quantum Ising model
the total geometric phase of the state is given by
\begin{eqnarray}
\Gamma_{final}(t=0)&=&\sum_{k, k\neq \pm \frac{\pi}{N}}\pi (1-\cos\theta_k)\\
& = & 2\pi \left( N-2+\cos\frac{\pi}{N}\right)
\end{eqnarray}

In a different case, if the system is driven slowly within the adiabatic limit and  ${\cal{N}}$ number of defects are produced, the geometric phase of the final state will be given by
\begin{eqnarray}
\Gamma_{final}(t=0)&=&\sum_{k,k \neq k_{0_i},i=1,2,...{\cal{N}}}\pi(1-\cos k)\\
&= & 2\pi(N-1)-2{\cal{N}}\pi +\pi\left(\cos \frac{{\cal{N}}\pi}{N}+\sin \frac{{\cal{N}}\pi}{N}\cot\frac{\pi}{2N}-1\right)
\end{eqnarray}
If $n$ be the average number of defects produced in the final state at $B=0$, then
\begin{equation}
{\cal{N}}= n N
\end{equation}
where $n \propto \frac{1}{\sqrt{\tau_q}}$.
Hence,
\begin{equation}
\Gamma
_{final}(t=0)=2\pi N(n-1)-3\pi +\pi\left(\cos\pi n +\sin \pi N\cot\frac{\pi}{2N}\right)
\end{equation}

Our analytical derivation shows that the in the adiabatic limit of a quench induced quantum phase transition, 
the geometric phase of the ground state in its final position with vanishing magnetic field depends on the average number of kinks, and hence on the quench time.  It is hoped that the present analysis may trigger investigations to explore the situation of physical realizations for obtaining the geometric phases associated with quench induced quantum phase transitions.  

To summarize, we have studied the dynamics of a quantum Ising model with a large number of spins and the geometric phase of the corresponding states. In our methodology we have adopted the novel idea of studying the dynamics of the geometric phase associated with a quantum phase transition when QPT is investigated in the light of {\it{KZ mechanism}}. The system initially at a high magnetic field was adiabatically evolved by gradually decreasing the magnetic field which is linearly adjustable by a time parameter $\tau_q$, called the quench time. When the magnetic field is turned off at $t=0$, the system, from its initial paramagnetic state, is finally in a ferromagnetic state with some defects generated in the system. The geometric phase of the final state depends on the average number of kinks produced in the system, which consequently depends on the quench time $\tau_q$. The present analysis traces the deformation of the geometric phase at criticality in a quench induced quantum phase transition associated with the generation of defects.


{\bf{Acknowlefgement}}: The author highly acknowledges the anonymous referees for their encouraging comments and constructive suggestions.

\newpage

\begin{figure}
\includegraphics[width=7cm]{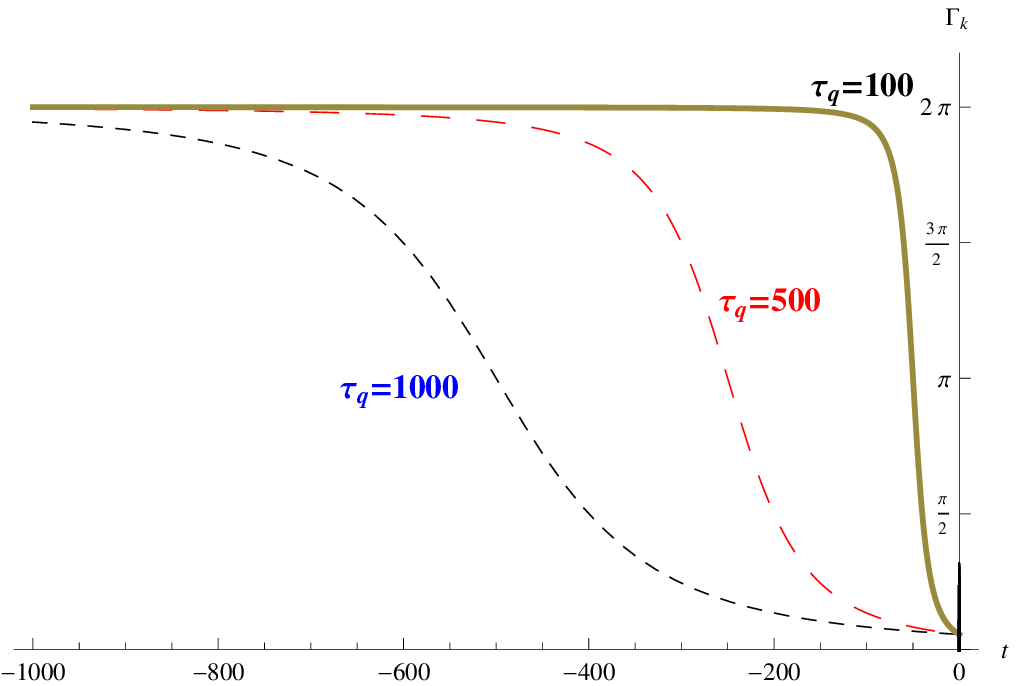}
\caption{Variation of $\Gamma_k$ with $t$ for fixed $\alpha=0.2$ and
$k=\pi/3$ with different $\tau_q$'s }
\end{figure}

\end{document}